\documentclass[conference, letterpaper]{IEEEtran}

\usepackage[letterpaper, top=0.76in, bottom=1.12in, left=0.68in, right=0.68in, columnsep=0.25in]{geometry}

\IEEEoverridecommandlockouts

\usepackage{cite}
\usepackage{amsmath,amssymb,amsfonts,amsthm} 
\usepackage{bm}

\usepackage[linesnumbered,ruled,vlined]{algorithm2e}

\usepackage{graphicx}
\usepackage{epstopdf}
\usepackage{textcomp}
\usepackage{xcolor}
\usepackage{array}
\usepackage{booktabs}
\usepackage{multirow}
\usepackage{caption}
\usepackage{subcaption} 

\usepackage[draft]{hyperref} 
\usepackage{xurl} 

\newcommand{\R}{\mathbb{R}}
\newcommand{\C}{\mathbb{C}}

\newcommand{\K}{\mathcal{K}}

\def\BibTeX{{\rm B\kern-.05em{\sc i\kern-.025em b}\kern-.08em
		T\kern-.1667em\lower.7ex\hbox{E}\kern-.125emX}}
\begin{document}

\title{6D Movable Antenna for Internet of Vehicles: CSI-Free Dynamic Antenna Configuration}



\author{\IEEEauthorblockN{Maoxin Ji$^{1,2}$, Qiong Wu$^{1,2,}$$^\ast{}$, Pingyi Fan$^{3}$, Kezhi Wang$^{4}$, Wen Chen$^{5}$ and Khaled B. Letaief$^{6}$}
     \IEEEauthorblockA{$^1$ School of Internet of Things Engineering, Jiangnan University, Wuxi 214122, China}
		\IEEEauthorblockA{$^2$ School of Information Engineering, Jiangxi Provincial Key Laboratory of Advanced Signal Processing \\and Intelligent Communications, Nanchang University, Nanchang 330031,China}
		\IEEEauthorblockA{$^3$ Department of Electronic Engineering, State Key laboratory of Space Network and Communications, \\Beijing National Research Center for Information Science and Technology, Tsinghua University, Beijing 100084, China}
\IEEEauthorblockA{$^4$ Department of Computer Science, Brunel University, London, Middlesex UB8 3PH, U.K}
\IEEEauthorblockA{$^5$ Department of Electronic Engineering, Shanghai Jiao Tong University, Shanghai 200240, China}
		\IEEEauthorblockA{$^6$ Department of Electrical and Computer Engineering, \\ the Hong Kong University of Science and Technology (HKUST), Hong Kong}
		{$^\ast{}$Corresponding author is Qiong Wu.}
	\IEEEauthorblockA{Email: maoxinji@stu.jiangnan.edu.cn, qiongwu@jiangnan.edu.cn,\\  
		fpy@tsinghua.edu.cn, Kezhi.Wang@brunel.ac.uk, wenchen@sjtu.edu.cn, eekhaled@ust.hk}

		\thanks{This work was supported in part by Jiangxi Province Science and Technology Development Programme under Grant No. 20242BCC32016, in part by the National Natural Science Foundation of China under Grant No. 61701197, 62531015, and U25A20399, in part by the Basic Research Program of Jiangsu under Grant BK20252084, in part by the National Key Research and Development Program of China under Grant No. 2021YFA1000500(4), in part by the Shanghai Kewei under Grant 24DP1500500, in part by the Hong Kong Research Grant Council under the Area of Excellence (AoE) Scheme Grant No. AoE/E-601/22-R and in part by the 111 Project under Grant No. B23008.}}
\maketitle

\begin{abstract}
Deploying six-dimensional movable antenna (6DMA) systems in Internet-of-Vehicles (IoV) scenarios can greatly enhance spectral efficiency. However, the high mobility of vehicles causes rapid spatio-temporal channel variations, posing a significant challenge to real-time 6DMA optimization. In this work, we pioneer the application of 6DMA in IoV and propose a low-complexity, instantaneous channel state information (CSI)-free dynamic configuration method. By integrating vehicle motion prediction with offline directional response priors, the proposed approach optimizes antenna positions and orientations at each reconfiguration epoch to maximize the average sum rate over a future time window. Simulation results in a typical urban intersection scenario demonstrate that the proposed 6DMA scheme significantly outperforms conventional fixed antenna arrays and simplified 6DMA baseline schemes in terms of total sum rate.
\end{abstract}


\section{Introduction} \label{Introduction}
The Internet of Vehicles (IoV) is experiencing rapidly growing communication demands, requiring ultra-low latency and ultra-high reliability\cite{7036784, 21, 25, 31}. However, the high mobility of vehicles introduces fast time-varying channel conditions, posing significant challenges to wireless network design \cite{survey, 23, 24, 32}. To accommodate the increasing demand, scaling up antenna deployments has become the mainstream trend. In IoV, multiple-input multiple-output (MIMO)\cite{mimo1} and even massive MIMO \cite{mimo2, 22} have been successively proposed. However, the continual growth in antenna scale leads to substantial hardware costs and markedly increased computational complexity in signal processing. Consequently, improving antenna utilization under a limited antenna budget is of critical importance.

Recently, Shao \textit{et al.} proposed a novel six-dimensional movable antenna (6DMA) technology  \cite{2}, which deploys antenna surfaces at different 3D positions with 3D rotations, providing highly flexible spatial distribution and achieving significant performance gains with a limited number of antennas. Moreover, 6DMA can adjust antenna positions by capturing slowly varying user distributions in space, resulting in a low adjustment frequency and offering clear advantages over existing fluid antenna system (FAS) \cite{3} and two-dimensional movable antenna technologies \cite{4}. 

However, this high deployment flexibility also introduces significant complexity in optimizing the positions and orientations of the 6DMA system. On one hand, the lack of efficient channel state information (CSI) estimation hinders quantitative assessment during antenna placement optimization. On the other hand, the joint 3D positioning and 3D rotation create a high-dimensional action space, making it difficult to obtain the global optimum. To address these challenges, reference \cite{1} proposed a Monte Carlo-based method to estimate average channel capacity and an alternating optimization scheme to obtain suboptimal antenna positions. In \cite{5}, a discrete 6DMA system with a finite set of feasible positions and rotations was constructed using the Fibonacci-sphere method under practical constraints, and an online optimization approach based on measured rates was developed. Reference \cite{7} introduced a fast CSI estimation method for 6DMA and a sequential optimization algorithm that first determines rotations and then selects positions. In \cite{8}, the directional sparsity of 6DMA systems was revealed, and the 6DMA positions and rotations were optimized by leveraging statistical CSI to estimate channel power. In addition, reference \cite{9} proposed a passive 6DMA scheme via an unmanned aerial vehicle (UAV)-mounted intelligent reflecting surface (IRS) to realize 3D position and orientation adjustments, along with an alternating optimization algorithm for the UAV’s position and attitude. These studies advanced 6DMA technology and verified its superiority over conventional antenna strategies.

Integrating 6DMA into IoV holds great promise, yet it entails significant technical challenges. On one hand, small-scale fading varies rapidly in IoV communications, and when combined with the complex 6DMA layout, predicting instantaneous CSI becomes extremely difficult, which impedes quantitative optimization of 6DMA positions and rotations. On the other hand, due to high vehicular mobility, user distributions change quickly. Therefore, 6DMA may need to adjust antenna positions frequently to attain near-instantaneous optimality or adjust more slowly to pursue long-term, robust suboptimal performance. Moreover, existing 6DMA optimization algorithms either estimate user rates via Monte Carlo methods \cite{1} or require extensive sampling over many antenna configurations \cite{5}, resulting in high complexity and limited efficiency for IoV scenarios.

To address these challenges, we formulate an optimization problem that maximizes the long-term sum rate in IoV environments. Vehicle locations are updated every time slot, while 6DMA positions are updated every $N$ slots. Based on predicted future vehicular distributions, antenna positions are adjusted to maximize the average sum rate over the next $N$ slots, without computing instantaneous CSI\footnote{The source code has been released at: \url{https://github.com/qiongwu86/6D-Movable-Antenna-for-Internet-of-Vehicles-CSI-Free-Dynamic-Antenna-Configuration}.}. Furthermore, leveraging the directional sparsity inherent to 6DMA antennas, we propose an online optimization scheme that exploits predicted side information and historical measured rate information to enable rapid allocation from historical knowledge. To the best of our knowledge, this is the first work to apply 6DMA to IoV scenarios.

\section{System Model} \label{system_model}
\begin{figure}[t]
	\centering
	\includegraphics[trim=0.5cm 0.5cm 0.5cm 0.5cm, clip, width=\columnwidth]{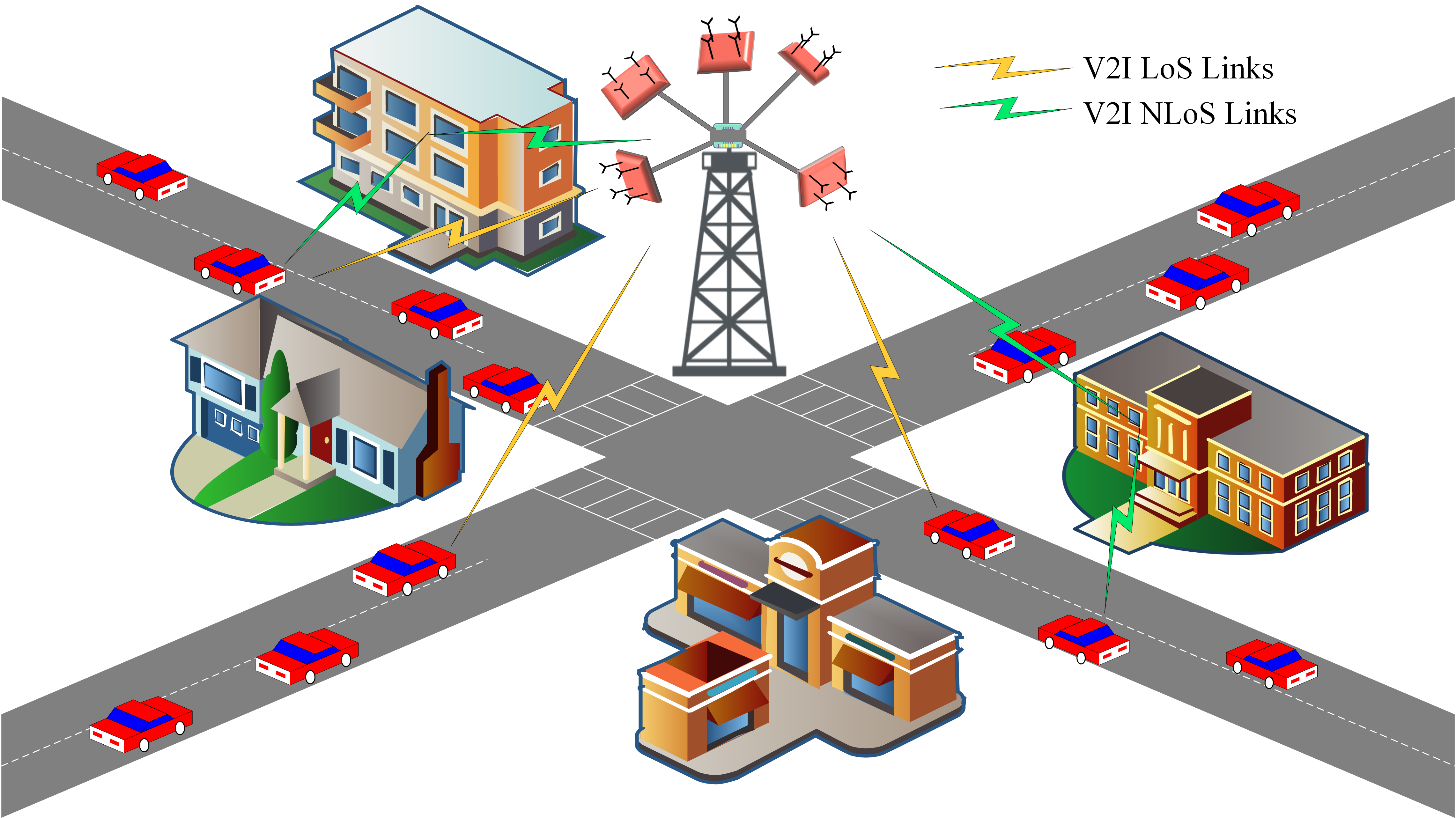}
	\caption{System Model}
	\label{system-model}
\end{figure}

We consider an urban cross-junction scenario as depicted in Fig.~\ref{system-model}, where vehicles are uniformly distributed along two perpendicular roads. Time is slotted with duration $\Delta t$, and vehicle positions are assumed quasi-static within each slot but updated at slot boundaries. At the beginning of every slot, the velocity of each vehicle is sampled from a truncated Gaussian distribution. Let $\K = \{1, 2, \ldots, K\}$ denote the set of vehicles. A single base station (BS) is located at the intersection center and serves all vehicles in the uplink. Each vehicle is equipped with a fixed-position antenna (FPA), while the BS employs a 6DMA system.

\subsection{Discrete-Position 6DMA Antenna System}

The 6DMA architecture consists of multiple antenna surfaces, each capable of independent 3D translation and rotation, connected to a central processing unit (CPU) via telescopic mechanical links. Due to hardware constraints, both position and orientation are restricted to discrete sets. Specifically, let $M$ and $J$ denote the numbers of feasible discrete positions and rotations, respectively. The catalog of admissible 6D configurations is defined as $\mathcal{P} = \{ (\mathbf{q}_i, \mathbf{r}_j) \mid i = 1,\ldots,M;\, j = 1,\ldots,J \},$ where $\mathbf{q}_i = [x_i, y_i, z_i]^\top \in \R^3$ is the center position of a surface and $\mathbf{r}_j = [\alpha_j, \beta_j, \gamma_j]^\top$ represents the Euler angles (in radians) that rotate the surface from its zero-rotation reference pose. In this reference pose, the outward unit normal vector aligns with $\mathbf{e}_x = [1, 0, 0]^\top$. For a given rotation $\mathbf{r}_j$, the resulting outward normal $\mathbf{n}_j$ is  
\begin{equation}
	\mathbf{n}_j =
	\begin{bmatrix}
		\cos\beta_j \cos\gamma_j \\
		\sin\alpha_j \sin\beta_j \cos\gamma_j - \cos\alpha_j \sin\gamma_j \\
		\cos\alpha_j \sin\beta_j \cos\gamma_j + \sin\alpha_j \sin\gamma_j
	\end{bmatrix}.
\end{equation}
Consider a deployment that chooses $U$ surfaces from the set $\mathcal{P}$, where $U \leq M$. To compactly represent the deployment configuration, we define a binary matrix $\mathbf{Z} \in \{0,1\}^{M \times J}$, where $[\mathbf{Z}]_{i,j} = 1$ if a surface is deployed at position $i$ with rotation $j$, and 0 otherwise. The total number of deployed surfaces is enforced by $\sum_{i=1}^{M} \sum_{j=1}^{N} [\mathbf{Z}]_{i,j} = U.$

The deployment should satisfy the following physical constraints. First, mutual reflection/blockage avoidance requires that for any two distinct deployed surfaces at $(i,j)$ and $(i',j')$ (i.e., $[\mathbf{Z}]_{i,j} = [\mathbf{Z}]_{i',j'} = 1$ and $(i,j) \neq (i',j')$), the surface at $(i,j)$ should not face the one at $(i',j')$, which is enforced by
\begin{equation}
	\mathbf{n}_j^\top (\mathbf{q}_{i'} - \mathbf{q}_i) \leq 0.
\end{equation}
Second, CPU blockage avoidance mandates that all deployed surfaces point away from the CPU, i.e.,
\begin{equation}
	\mathbf{n}_j^\top \mathbf{q}_i \geq 0,
\end{equation}
for all $(i,j)$ such that $[\mathbf{Z}]_{i,j} = 1$. Third, minimum inter-surface spacing ensures that any two surfaces at distinct positions are separated by at least $d_{\min}$, i.e.,
\begin{equation}
	\|\mathbf{q}_i - \mathbf{q}_{i'}\|_2 \geq d_{\min},
\end{equation}
for all $(i,j), (i',j')$ with $[\mathbf{Z}]_{i,j} = [\mathbf{Z}]_{i',j'} = 1$ and $i \neq i'$. Finally, one surface per position is imposed by
\begin{equation}
	\sum_{j=1}^{J} [\mathbf{Z}]_{i,j} \leq 1, \quad \forall\, i = 1,\ldots,M.
\end{equation}

\subsection{Channel Model}

We consider the uplink transmission from $K$ vehicles to the BS. Each 6DMA surface is equipped with $Q$ FPA elements, yielding a total of $U*Q$ receive antennas. We adopt a hybrid near--far field model \cite{10}: propagation between different surfaces is in the far-field regime, while intra-surface element responses account for near-field spherical wavefronts. Let $\mathbf{p}_k \in \R^3$ denote the position of vehicle $k$. The direction from vehicle $k$ to surface at $(i,j)$ is  
\begin{equation}
	\hat{\mathbf{d}}_{k,i,j} = \frac{\mathbf{p}_k - \mathbf{q}_i}{\|\mathbf{p}_k - \mathbf{q}_i\|}.
\end{equation}
To characterize the local incident angles, we construct an orthonormal basis on the tangent plane of surface $(i,j)$. With $\mathbf{e}_{\mathrm{ref}} = [0, 0, 1]^\top$, define  
	$\mathbf{u}_{i,j} = \frac{\mathbf{n}_j \times \mathbf{e}_{\mathrm{ref}}}{\|\mathbf{n}_j \times \mathbf{e}_{\mathrm{ref}}\|}, \quad
	\mathbf{v}_{i,j} = \mathbf{n}_j \times \mathbf{u}_{i,j},$
and the tangential projection $\hat{\mathbf{d}}^{\parallel}_{k,i,j} = 
	\frac{\hat{\mathbf{d}}_{k,i,j} - (\mathbf{n}_j^\top \hat{\mathbf{d}}_{k,i,j}) \mathbf{n}_j}
	{\|\hat{\mathbf{d}}_{k,i,j} - (\mathbf{n}_j^\top \hat{\mathbf{d}}_{k,i,j}) \mathbf{n}_j\|}.$
The local elevation and azimuth angles are then  
\begin{equation}
	\tilde{\theta}_{k,i,j} = \arccos(\mathbf{n}_j^\top \hat{\mathbf{d}}_{k,i,j}),
\end{equation}
\begin{equation}
	\tilde{\phi}_{k,i,j} = \operatorname{atan2}\big( \hat{\mathbf{d}}^{\parallel\top}_{k,i,j} \mathbf{v}_{i,j},\, \hat{\mathbf{d}}^{\parallel\top}_{k,i,j} \mathbf{u}_{i,j} \big).
\end{equation}

Following the 3GPP radiation pattern model \cite{15}, the horizontal and vertical attenuation terms are  
\begin{equation}
	\left\{
	\begin{aligned}
		A_H &= -\min\!\left(12\left(\frac{\tilde{\phi}_{k,i,j}}{\phi_{3\mathrm{dB}}}\right)^2,\, A_m\right), \\
		A_V &= -\min\!\left(12\left(\frac{\tilde{\theta}_{k,i,j}}{\theta_{3\mathrm{dB}}}\right)^2,\, A_m\right),
	\end{aligned}
	\right.
\end{equation}
and the linear-scale element gain is $G^{(\mathrm{lin})}_{k,i,j} = 10^{\frac{G_{\max} - \min( -(A_H + A_V), A_m )}{10}}$,  where $\phi_{3\mathrm{dB}}$ and $\theta_{3\mathrm{dB}}$ are the horizontal and vertical half-power beamwidths, respectively, $A_m$ denotes the maximum sidelobe attenuation, and $G_{\max}$ is the maximum antenna gain in the boresight direction.

For each potential path $\ell$ between vehicle $k$ and surface $(i,j)$, the path loss $\mathrm{PL}_{k,i,j,\ell}$ is determined by the 3GPP urban microcell (UMi) model \cite{15} based on the geometric distance $d_{k,i} = \|\mathbf{p}_k - \mathbf{q}_i\|$. The Line-of-Sight (LoS) probability is  
\begin{equation}
	p_{\mathrm{LoS}}(d_{k,i}) = \min\!\left(\frac{d_1}{d_{k,i}}, 1\right)\big(1 - e^{-d_{k,i}/d_2}\big) + e^{-d_{k,i}/d_2},
\end{equation}
where $d_1=18$, $d_2=36$. For each path $\ell$, a Monte Carlo trial is performed: With probability $p_{\mathrm{LoS}}(d_{k,i})$, the link is LoS and  
\begin{equation}
	\mathrm{PL}_{k,i,j,\ell} = 32.4 + 21\log_{10} d_{k,i} + 20\log_{10} f_c,
\end{equation}
otherwise it is Non-Line-of-Sight (NLoS) and  
\begin{equation}
	\mathrm{PL}_{k,i,j,\ell} = 35.3 \log_{10} d_{k,i}+ 22.4 + 21.3\log_{10} f_c + \chi_\ell,
\end{equation}
where $\chi_\ell \sim \mathcal{N}(0, \sigma_{SF}^2)$ and $\sigma_{SF} = 7.82~\text{dB}$. The corresponding linear large-scale power gain is defined as $\eta^{(\mathrm{LS})}_{k,i,j,\ell} = 10^{-\mathrm{PL}_{k,i,j,\ell}/10}$.

Let the global coordinates of the $m$-th element on surface $(i,j)$ be  
\begin{equation}
	\mathbf{c}_{i,j,m} = \mathbf{q}_i + u_m \mathbf{u}_{i,j} + v_m \mathbf{v}_{i,j},
\end{equation}
with $u_m, v_m \in \{-s/2, s/2\}$ for a square array of side length $s$. The near-field spherical-wave phase from vehicle $k$ to this element is  
\begin{equation}
	\phi_{k,i,j,m,\ell} = -\frac{2\pi}{\lambda} \|\mathbf{p}_k - \mathbf{c}_{i,j,m}\|.
\end{equation}
Assuming a small set of dominant paths $\ell \in \mathcal{D}_{k,i,j}$, each with independent Rayleigh fading $\xi_{k,i,j,\ell} \sim \mathcal{CN}(0,1)$, the per-element channel coefficient is  
\begin{equation}
	h_{k,i,j,m} = \sum_{\ell \in \mathcal{D}_{k,i,j}} 
	\sqrt{\eta^{(\mathrm{LS})}_{k,i,j,\ell}} \cdot 
	\sqrt{G^{(\mathrm{lin})}_{k,i,j,\ell}} \cdot 
	e^{j \phi_{k,i,j,m,\ell}} \cdot 
	\xi_{k,i,j,\ell},
\end{equation}
where the near-field effect is explicitly captured by the element-dependent phase term $e^{j \phi_{k,i,j,m,\ell}}$, while the directional radiation pattern and large-scale path loss are modeled via $G^{(\mathrm{lin})}_{k,i,j,\ell}$ and $\eta^{(\mathrm{LS})}_{k,i,j,\ell}$, respectively.

Stacking across elements and surfaces, the channel vector for user $k$ is $\mathbf{h}_k = [\cdots, \mathbf{h}_{k,i,j}^\top, \cdots]^\top \in \C^{BQ}$, where $\mathbf{h}_{k,i,j} = [h_{k,i,j,1}, \ldots, h_{k,i,j,Q}]^\top$ for each $(i,j)$ with $[\mathbf{Z}]_{i,j}=1$, and the overall channel matrix is $\mathbf{H} = [\mathbf{h}_1, \ldots, \mathbf{h}_K] \in \C^{BQ \times K}$. The computation of \(\mathbf{H}(t)\) can be regarded as a mapping from the vehicle positions \(\{\mathbf{p}_k(t)\}_{k=1}^K\) and the 6DMA configuration \(\mathbf{Z}(t)\), expressed as: $\mathbf{H}(t) = \mathcal{H}\big( \{ \mathbf{p}_k(t) \}_{k=1}^{K},\, \mathbf{Z}(t) \big).$

The matched-filter signal-to-interference-plus-noise ratio (SINR) \cite{28, 29, 30} for user $k$ is given by
\begin{equation}
	\Gamma_k\big( \mathbf{H}\big) = \frac{P_k \|\mathbf{h}_k\|^2}
	{\sum_{j \neq k} P_j \frac{|\mathbf{h}_k^{\mathrm{H}} \mathbf{h}_j|^2}{\|\mathbf{h}_j\|^2 + \varepsilon} + \sigma^2},
	\label{eq:sinr_model}
\end{equation}
with a regularization constant $\varepsilon > 0$, where $\sigma^2$ denotes the noise power at the receiver. The achievable rate is  
\begin{equation}
	R_k = B \log_2(1 + \Gamma_k\big( \mathbf{H} \big)),
\end{equation}
and the sum rate is $R_{\mathrm{sum}} = \sum_{k=1}^{K} R_k,$ where $B$ denotes the system bandwidth.

\subsection{Problem Formulation}

Due to the high mobility of vehicles, user positions evolve rapidly, resulting in fast-changing spatial distributions. While the 6DMA system can enhance performance by adapting its configuration to the user distribution, mechanical and computational constraints prohibit per-slot reconfiguration.

We discretize time into slots of duration $\Delta t$. The position of vehicle $k$ at slot $t$, denoted $\mathbf{p}_k(t) \in \mathbb{R}^3$, follows the linear kinematic model
\begin{equation}
	\mathbf{p}_k(t+1) = \mathbf{p}_k(t) + v_k(t) \, \mathbf{a}_k(t) \, \Delta t,
\end{equation}
where $v_k(t) \geq 0$ is the instantaneous speed and $\mathbf{a}_k(t) \in \mathbb{R}^3$ is the unit direction vector.

To balance adaptivity and practicality, we update $\mathbf{Z}$ once every $N$ slots. During the $l$-th optimization cycle, covering slots $t \in \{(l-1)N+1, \ldots, lN\}$, the configuration remains fixed: $\mathbf{Z}(t) = \mathbf{Z}_l$.

Leveraging the spatial-temporal regularity of vehicular traffic, we predict future positions using the instantaneous average velocity $\bar{v}(t) = \frac{1}{K} \sum_{k=1}^{K} v_k(t)$, and recursively compute $\hat{\mathbf{p}}_k(t+1) = \hat{\mathbf{p}}_k(t) + \mathbf{a}_k(t) \bar{v}(t) \Delta t$, with $\hat{\mathbf{p}}_k(0) = \mathbf{p}_k(0)$.

The average total user rate over the next \(N\) time slots based on the predicted positions can be computed as follows. Let \( t_l \triangleq (l-1)N+1 \) denote the starting time index of the \(l\)th period. Then,
\begin{equation}
	C_{\mathrm{avg}}(\mathbf{Z}_l) \triangleq \sum_{t=t_l}^{t_l+N-1} \sum_{k=1}^{K} \frac{B}{N} \log_2\!\Big( 1 + \Gamma_k\big( \mathbf{H} \Big).
\end{equation}
The complete optimization problem can be formulated as follows:
\allowdisplaybreaks
\begin{subequations}
	\begin{align}
		\max_{\mathbf{Z}_l}\quad & C_{\mathrm{avg}}(\mathbf{Z}_l) \label{eq:P1_obj} \\
		\text{s.t.} \quad 
		& \sum_{i=1}^{M} \sum_{j=1}^{N} [\mathbf{Z}_l]_{i,j} = U, \label{eq:P1_total} \\
		& \sum_{j=1}^{N} [\mathbf{Z}_l]_{i,j} \leq 1, \quad \forall\, i = 1,\ldots,M, \label{eq:P1_one_per_pos} \\
		& \mathbf{n}_j^\top (\mathbf{q}_{i'} - \mathbf{q}_i) \leq 0, \quad \forall\, (i,j), (i',j') , \label{eq:P1_mutual} \\
		& \mathbf{n}_j^\top \mathbf{q}_i \geq 0, \quad \forall\, (i,j), \label{eq:P1_cpu} \\
		& \|\mathbf{q}_i - \mathbf{q}_{i'}\|_2 \geq d_{\min}, \quad \forall\, (i,j), (i',j') \label{eq:P1_spacing} \\
		& [\mathbf{Z}_l]_{i,j} \in \{0,1\}, \quad \forall\, (i,j), \label{eq:P1_binary}
	\end{align}
\end{subequations}
where constraint \eqref{eq:P1_total} ensures exactly $U$ surfaces are deployed. Constraint \eqref{eq:P1_one_per_pos} enforces that each discrete position hosts at most one surface. Constraints \eqref{eq:P1_mutual}–\eqref{eq:P1_spacing} encode the physical feasibility requirements: mutual reflection/blockage avoidance, CPU visibility, and minimum inter-surface spacing, respectively. Constraint \eqref{eq:P1_binary} imposes binary selection over the discrete configuration catalog $\mathbf{Z}_l$.

This formulation enables proactive, low-complexity 6DMA reconfiguration that exploits spatial-temporal predictability in vehicular traffic, while respecting hardware limitations and avoiding reliance on instantaneous channel state information.

\section{Optimization Method}
In this section, we introduce a grid-based method for determining feasible locations and orientations. Furthermore, we exploit 6DMA sparsity in an online optimization fusing pre-test and historical rates.

\subsection{Discrete Position and Orientation Construction}

To address the lack of explicit neighborhood structure in random spherical discretization schemes (e.g., Fibonacci sampling) \cite{26, 27}, we employ a deterministic grid on a sphere of radius $R$ around the BS. The sphere is divided into $F$ equally spaced meridians, yielding the azimuth set $\Phi = \left\{ \varphi_f = \frac{2\pi f}{F} \right\}_{f=0}^{F-1}$.

To satisfy the minimum inter-surface distance constraint $d_{\min}$, we first determine the first valid latitude circle near each pole. Let the chord length between adjacent points on this circle (along meridians) be at least $d_{\min}$. The corresponding circle radius should satisfy $r_{\text{first}} = \frac{d_{\min}}{2\sin(\pi / F)}$.
The polar angle is then $\theta_{\text{first}} = \arccos(r_{\text{first}} / {r_0})$, where $r_0$ denotes the radius of the sphere. The axial distance from the pole to this latitude circle is $d_{\text{pole}} = r_0 (1 - \cos\theta_{\text{first}})$. The available axial length $2{r_0} - 2d_{\text{pole}}$ yields the maximum number of intermediate latitude circles:
\begin{equation}
	L = \left\lfloor \frac{2{r_0} - 2d_{\text{pole}}}{d_{\min}} \right\rfloor,
\end{equation}
where $\lfloor \cdot \rfloor$ denotes the floor function.
Including the two poles, the total number of discrete positions is $M = F \cdot L + 2$, where each non-polar position is represented in spherical coordinates as $({r_0}, \theta_g, \varphi_f)$, where $\theta_g$ denotes the polar angle of the $g$-th intermediate latitude circle.

To define orientation degrees of freedom, an 8-neighborhood structure is established for each non-polar position, comprising adjacent points along the same meridian (up/down), same latitude (left/right), and four diagonal neighbors. For any two adjacent neighbors $\mathbf{q}_{i_1}, \mathbf{q}_{i_2}$ and the current position $\mathbf{q}_i$, the unit normal of the triangle they form is
\vspace{-0.2cm}
\begin{equation}
	\mathbf{n}_i^{(m)} = \frac{(\mathbf{q}_{i_1} - \mathbf{q}_i) \times (\mathbf{q}_{i_2} - \mathbf{q}_i)}{\|(\mathbf{q}_{i_1} - \mathbf{q}_i) \times (\mathbf{q}_{i_2} - \mathbf{q}_i)\|}, \quad m = 1, \ldots, 8.
\end{equation}
Additionally, the radial outward normal is included: $\mathbf{n}_i^{(0)} = \frac{\mathbf{q}_i}{\|\mathbf{q}_i\|}$, yielding $J = 9$ discrete orientations per position. The poles are handled via virtual neighbors to also maintain 9 orientations. The total number of feasible configurations is $|\mathcal{P}| = M \times J$. This construction inherently satisfies the minimum spacing constraint $\|\mathbf{q}_i - \mathbf{q}_{i'}\|_2 \geq d_{\min}$ and the CPU blockage avoidance condition $\mathbf{n}_j^\top \mathbf{q}_i \geq 0$.

\subsection{Offline Response Modeling}

The 6DMA system exhibits strong directional sparsity: a single surface provides significant gain only to users in a limited spatial region. To quantify this, a one-time offline profiling is performed prior to deployment to establish a ``spatial region--preferred configuration'' prior mapping.

The ground service area $[0, X] \times [0, Y]$ is partitioned into square grids of side length $W$, resulting in a total of $N_{\text{grid}} = \left\lceil \frac{X}{W} \right\rceil \cdot \left\lceil \frac{Y}{W} \right\rceil$ grids, where $\lceil \cdot \rceil$ denotes the ceiling function. The center of grid $(g_x, g_y)$ is
\begin{equation}
	\mathbf{c}_{g_x,g_y} = \left( (g_x - \tfrac{1}{2})W,\ (g_y - \tfrac{1}{2})W,\ z_{\text{veh}} \right)^\top,
\end{equation}
where $z_{\text{veh}}$ denotes the typical vehicle height.

Within each grid, $S$ representative user positions $\{\mathbf{p}_g^{(s)}\}_{s=1}^S$ are uniformly sampled. Using a single 6DMA surface (equipped with a $2 \times 2$ FPA array) as the radiation unit, the average theoretical rate of each feasible configuration $(\mathbf{q}_i, \mathbf{r}_j) \in \mathcal{P}$ for the grid is evaluated as
\begin{equation}
	\bar{r}_g(i,j) = \frac{1}{S} \sum_{s=1}^{S} B \log_2 \left( 1 + \mathrm{SINR}\big( \mathbf{p}_g^{(s)}; \mathbf{q}_i, \mathbf{r}_j \big) \right),
\end{equation}
where the SINR is given by \eqref{eq:sinr_model}.

For each grid, the top-$H$ configurations form the offline candidate set:
\begin{equation}
	\mathcal{C}_g = \left\{ (\mathbf{q}_i, \mathbf{r}_j) \in \mathcal{P} \,\middle|\, \bar{r}_g(i,j) \text{ ranks among the top } H \right\}.
\end{equation}
This prior library captures the directional response characteristics under stable scattering and macroscopic geometry, providing a high-quality, compact candidate pool for online decision-making.

\subsection{Online Optimization Based on Historical Rates}

Since dynamic vehicle distributions cause deviations from offline assumptions, we introduce an online mechanism to refine priors using real-time rates.

Let the $l$-th reconfiguration cycle span time slots $\mathcal{T}_l$, with 6DMA configuration $\mathbf{Z}_{l}$ and vehicle positions $\{\mathbf{p}_k(t)\}$. The average rate for grid $g$ in this cycle is
\begin{equation}
	\hat{R}_g^{l} = \frac{1}{|\mathcal{K}_g^{l}| \cdot |\mathcal{T}_l|} \sum_{k \in \mathcal{K}_g^{l}} \sum_{t \in \mathcal{T}_l} R_k(t),
	\label{eq:avg_rate}
	\vspace{-0.1cm}
\end{equation}
where $\mathcal{K}_g^{l}$ is the set of users mapped to grid $g$. The rate change $\Delta R_g^{l} = \hat{R}_g^{l} - \hat{R}_g^{l-1}$ drives prior updates.

Each grid maintains a dynamic candidate set $\mathcal{S}_g$, where each configuration $(\mathbf{q}_i, \mathbf{r}_j)$ is associated with a quality score $q_{g,i,j} \in [0,1]$ and a usage count $\nu_{g,i,j}$. The update rules are as follows. When configuration $(\mathbf{q}_i, \mathbf{r}_j)$ is active in cycle $l$, its usage count is always incremented ($\nu_{g,i,j} \gets \nu_{g,i,j} + 1$), and its quality score $q_{g,i,j}$ is adjusted based on the observed rate change $\Delta R_g^l$: if $\Delta R_g^l > \delta$, then $q_{g,i,j}$ is increased by $\eta_+ \Delta R_g^l$; if $\Delta R_g^l < -\delta$, it is decreased by $\eta_- |\Delta R_g^l|$; otherwise, it remains unchanged. The score is always projected onto the interval $[0,1]$, where $\eta_+ > 0$ and $\eta_- > 0$ are step sizes that control the adaptation speed for positive and negative performance feedback, respectively, and $\delta > 0$ is a sensitivity threshold that filters out negligible rate variations. New high-performing configurations are admitted with an initial score $q_0 \in (0,1]$, and the lowest-scoring entry is removed upon capacity overflow.

At each reconfiguration instant, the candidate sets of all active grids (i.e., $|\mathcal{K}_g^{l}| > 0$) are aggregated into a global candidate set $\mathcal{U} = \bigcup_g \mathcal{S}_g$. For each candidate $(\mathbf{q}_i, \mathbf{r}_j) \in \mathcal{U}$, a priority score is computed as
\begin{equation}
	\pi_{g,i,j} = q_{g,i,j} \cdot \rho_g \cdot w_{\text{rank}} \cdot (1 + \kappa \cdot \nu_{g,i,j}),
	\label{eq:priority_score}
\end{equation}
where $\rho_g$ is the grid demand weight (e.g., the current number of users in grid $g$), $w_{\text{rank}}$ is a rank-based weighting factor (e.g., $w_{\text{rank}} = H - \mathrm{rank} + 1$ with $H$ denoting the maximum rank), and $\kappa > 0$ is an exploration bonus coefficient that promotes frequently used configurations.

A greedy selection strategy picks configurations in descending order of $\pi_{g,i,j}$ until $U$ surfaces are allocated. A set of selected positions $\mathcal{Q}_{\text{sel}}$ is maintained to ensure that no duplicate antenna positions are selected.

\subsection{Prediction-Enhanced Performance Optimization}

To improve long-term performance under low-frequency reconfiguration, a motion prediction mechanism is employed. At each reconfiguration instant, future vehicle positions $\{\hat{\mathbf{p}}_k(t)\}_{t=1}^{N}$ over the next $N$ slots are predicted using a motion model. The cumulative demand density for grid $g$ is then
\begin{equation}
	\tilde{\rho}_g = \sum_{t=1}^{N} \sum_{k=1}^{K} \mathbf{1}\big( \hat{\mathbf{p}}_k(t) \in \text{grid } g \big).
\end{equation}
Using $\tilde{\rho}_g$ as the demand weight $\rho_g$ in priority computation shifts the optimization objective from instantaneous to time-averaged performance over a prediction window, significantly enhancing robustness.

Complexity Analysis: The proposed method ensures low latency by avoiding real-time CSI estimation. The complexity is dominated by sorting the priority scores of offline candidates. With $K$ vehicles and $H$ candidates per active grid, the search space scales as $\mathcal{O}(KH)$. Consequently, the sorting complexity is $\mathcal{O}(KH \log (KH))$, which is computationally negligible compared to conventional iterative optimization algorithms or exhaustive search strategies, making it highly suitable for high-mobility IoV scenarios.


\begin{figure*}[t]
	\centering
	\begin{minipage}[b]{0.28\textwidth}
		\centering
		\includegraphics[width=\textwidth]{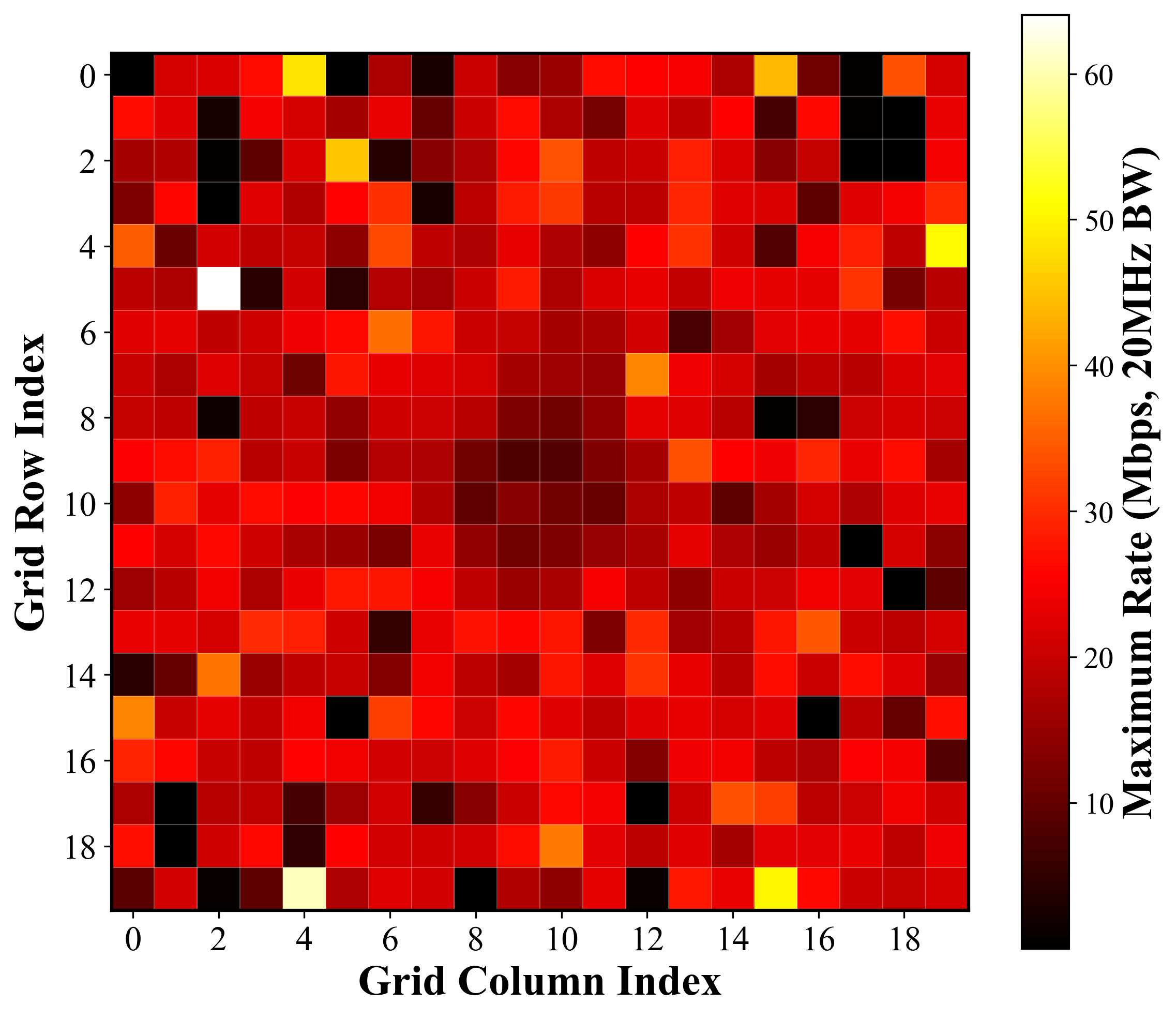}
		\caption{Grid Heatmap}
		\label{fig2}
	\end{minipage}
	\hfill
	\begin{minipage}[b]{0.32\textwidth}
		\centering
		\includegraphics[width=\textwidth]{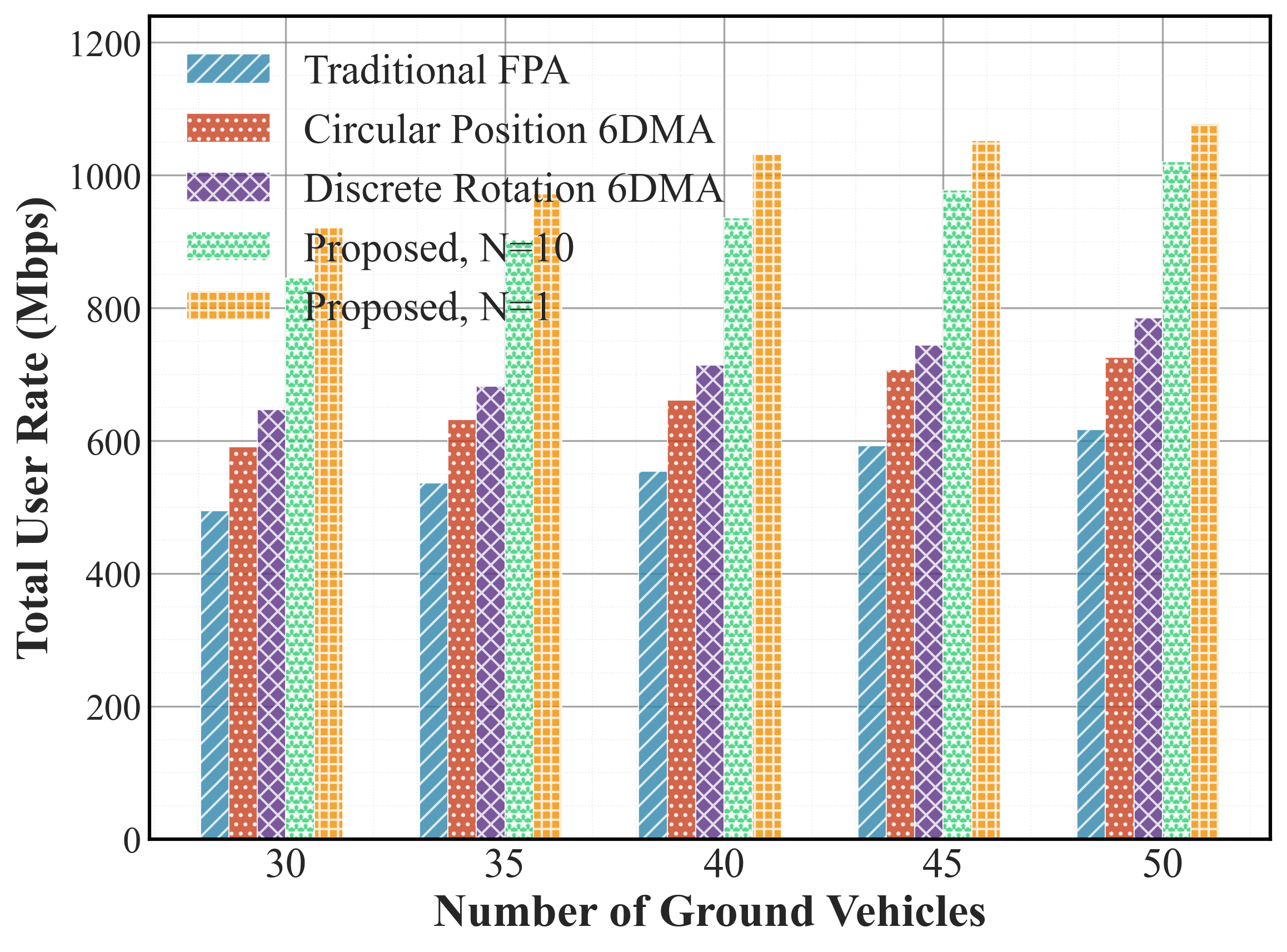}
		\caption{Total Rate Versus User Count}
		\label{fig3}
	\end{minipage}
	\hfill
	\begin{minipage}[b]{0.32\textwidth}
		\centering
		\includegraphics[width=\textwidth]{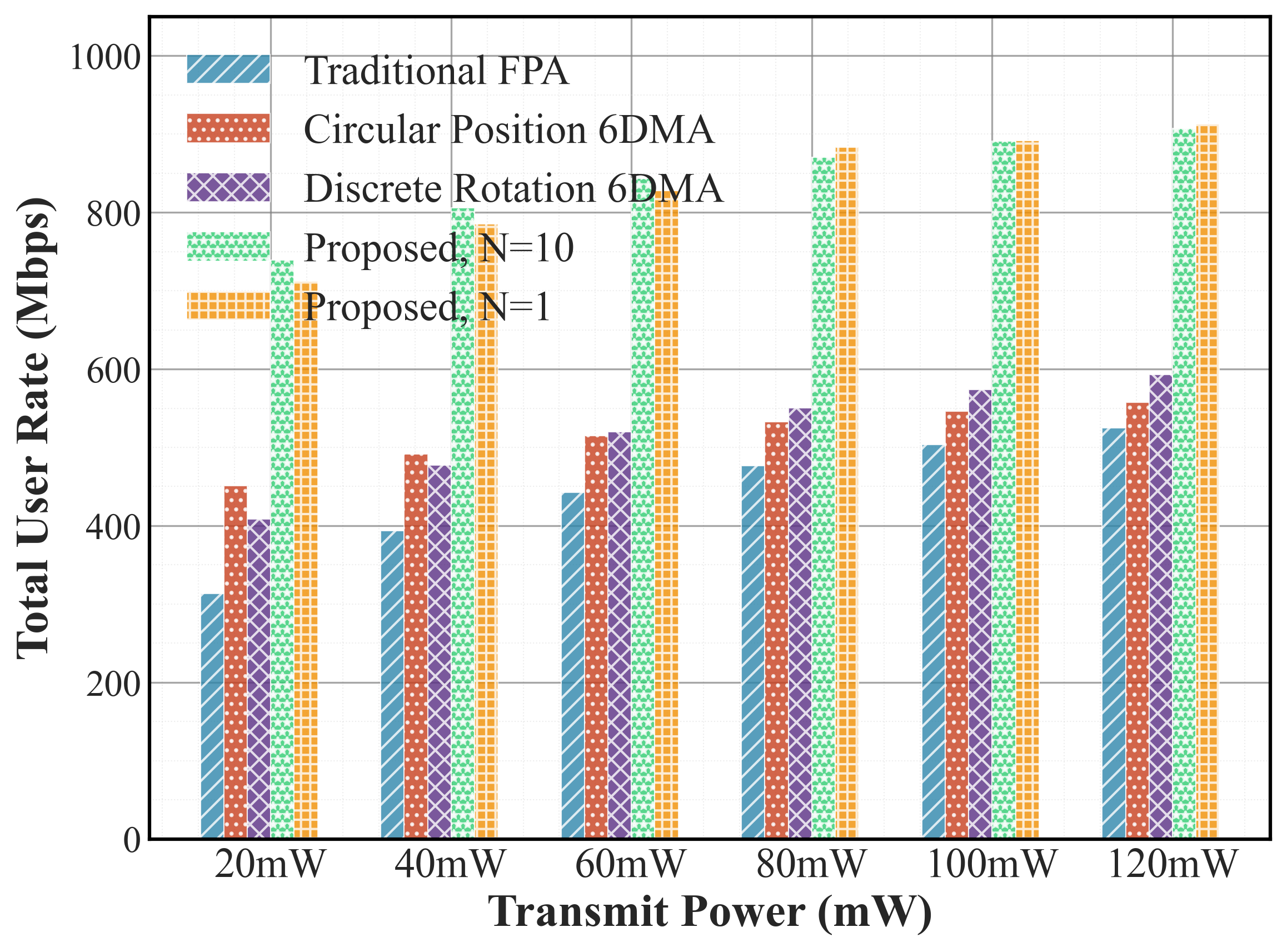}
		\caption{Total Rate Versus Power}
		\label{fig4}
	\end{minipage}
	\vspace{-0.2cm}
\end{figure*}
\begin{table}\footnotesize
\caption{Values of parameters}
\label{tab1}
\centering
\begin{tabular}{|c|c|c|c|}
\hline
\textbf{Parameter} &\textbf{Value} &\textbf{Parameter} &\textbf{Value}\\
\hline
$K$ & 30/35/40/45/50 & $v$ & $10-20m/s$ \\
\hline
$U$ & $16$ & $Q$ & $4(2*2)$ \\
\hline
$B$ & $20MHz$ & $N$ & $10/1$ \\
\hline
$d_{min}$ & $10cm$ & $R$ & $0.5m$ \\
\hline
$X*Y$ & $300m*300m$ & $W$ & $15m$ \\
\hline
$F$ & $12$ & $S$ & $20$ \\
\hline
\end{tabular}
\end{table}

\section{Simulation Results}

In this section, we present simulation results based on Python~3.8 to validate the performance superiority of the proposed 6DMA antenna system in IoV scenarios. The detailed simulation parameters are summarized in Table~\ref{tab1}. We compare against three baseline schemes: (i) a conventional FPA system with four $90^\circ$ sectors, each equipped with a fixed $4 \times 4$ rectangular antenna array and a uniform $15^\circ$ downtilt; (ii) a circular-discrete-position 6DMA, where four sectorized surfaces move along a ground-parallel circular trajectory; and (iii) 6DMA with discrete rotations only, in which the surface positions are fixed but each can rotate within a predefined angular range using discrete orientation steps.

Fig.~\ref{fig2} illustrates the average per-user rate achievable by a single optimally configured 6DMA surface for each ground grid cell. Users near the base station (center region) exhibit relatively low rates, primarily because their incident angles to the antenna surface are often large, resulting in low antenna gain. In contrast, users in the surrounding annular region benefit from both low path loss and favorable incident angles, yielding the highest rates. Users in the four corner areas suffer from severe blockage and long propagation distances, leading to the worst performance.

Fig.~\ref{fig3} shows the total sum rate versus the number of vehicles, where $N$ denotes the reconfiguration interval (in time slots) of the 6DMA surfaces. As expected, the sum rate increases with the number of users. The proposed 6DMA system significantly outperforms all baselines due to its flexible spatial deployment. Although the strategy that reconfigures every slot ($N=1$) achieves the highest performance, the predictive scheme with $N=10$ (i.e., reconfiguration every 10 slots) incurs only a minor performance loss while being far more practical in real-world deployments.

Fig.~\ref{fig4} depicts the sum rate versus transmit power. Interestingly, at low transmit power levels, the $N=10$ scheme slightly outperforms the $N=1$ scheme, and the circular-position 6DMA also marginally surpasses the discrete-orientation-only variant. This can be attributed to the fact that under low signal-to-noise ratio conditions, system performance is more sensitive to coverage uniformity and beam stability than to reconfiguration frequency. Coarser reconfiguration (e.g., $N=10$) mitigates the adverse effects of prediction errors and channel fluctuations, thereby offering more robust performance in power-limited regimes. Moreover, circular positional adjustment provides greater spatial degrees of freedom compared to orientation-only tuning, enabling more effective path loss compensation when transmit power is constrained. As the transmit power increases and channel conditions improve, the advantage of frequent reconfiguration becomes increasingly pronounced.
\vspace{-0.1cm}
\section{Conclusions}
In this paper, we explore the application of the 6DMA system in highly dynamic IoV scenarios. To address the challenges posed by rapidly varying channel conditions and user distributions due to high vehicle mobility, we propose a low-complexity, CSI-free optimization method that leverages predicted vehicle distributions to adaptively configure the 6DMA system and maximize the total user sum rate. Simulation results demonstrate that the proposed 6DMA system significantly outperforms conventional antenna deployment strategies, owing to its flexible spatial reconfigurability. Future work will focus on refining antenna placement at a finer granularity and minimizing the number of physically moved antenna surfaces during reconfiguration to reduce mechanical overhead and energy consumption.


\begin{thebibliography}{99}
	\bibitem{7036784}
	Q.~Wu and J.~Zheng, ``Performance modeling of the IEEE 802.11p EDCA mechanism for VANET,'' in \emph{Proc. IEEE Global Commun. Conf. (GLOBECOM)}, Austin, TX, USA, Dec. 2014, pp. 57--63.
	
	\bibitem{21}
	Q.~Wu, X.~Wang, Q.~Fan, P.~Fan, C.~Zhang, and Z.~Li, ``High stable and accurate vehicle selection scheme based on federated edge learning in vehicular networks,'' \emph{China Commun.}, vol.~20, no.~3, pp. 1--17, 2023.
	
	\bibitem{25}
	Q.~Wang, D.~O. Wu, and P.~Fan, ``Delay-constrained optimal link scheduling in wireless sensor networks,'' \emph{IEEE Trans. Veh. Technol.}, vol.~59, no.~9, pp. 4564--4577, 2010.
	
	\bibitem{31}
	Q.~Wu and J.~Zheng, ``Performance modeling of the IEEE 802.11 p EDCA mechanism for VANET,'' in \emph{Proc. IEEE Global Commun. Conf. (GLOBECOM)}, Austin, TX, USA, pp. 57--63, 2014.

	\bibitem{survey}
	C.~Li, M.~Dong, Y.~Fu, F.~R.~Yu, and N.~Cheng, ``Integrated sensing, communication, and computation for IoV: Challenges and opportunities,'' \emph{IEEE Commun. Surveys Tuts.}, early access, 2025.
	
	\bibitem{23}
	Y.~Yang and P.~Fan, ``Doppler frequency offset estimation and diversity reception scheme of high-speed railway with multiple antennas on separated carriage,'' \emph{J. Mod. Transp.}, vol.~20, no.~4, pp. 227--233, 2012.
	
	\bibitem{24}
	H.~Zhou, P.~Fan, and J.~Li, ``Global proportional fair scheduling for networks with multiple base stations,'' \emph{IEEE Trans. Veh. Technol.}, vol.~60, no.~4, pp. 1867--1879, 2011.
	
	\bibitem{32}
	J.~Fan, S.~Yin, Q.~Wu, and F.~Gao, ``Study on refined deployment of wireless mesh sensor network,'' in \emph{Proc. 6th Int. Conf. Wireless Commun. Netw. Mobile Comput. (WiCOM)}, Chengdu, China, 2010.
	
	\bibitem{mimo1}
	J.-H.~Jo, J.-N.~Shim, B.~Kim, C.-B.~Chae, and D.~K.~Kim, ``AoA-based position and orientation estimation using lens MIMO in cooperative vehicle-to-vehicle systems,'' \emph{IEEE J. Sel. Areas Commun.}, vol.~41, no.~12, pp. 3719--3735, Dec. 2023.
	
	\bibitem{mimo2}
	H.~Jiang, Z.~Zhang, J.~Dang, and L.~Wu, ``A novel 3-D massive MIMO channel model for vehicle-to-vehicle communication environments,'' \emph{IEEE Trans. Commun.}, vol.~66, no.~1, pp. 79--90, Jan. 2018.
	
	\bibitem{22}
	X.~Chen, J.~Lu, P.~Fan, and K.~B. Letaief, ``Massive MIMO beamforming with transmit diversity for high mobility wireless communications,'' \emph{IEEE Access}, vol.~5, pp. 23032--23045, 2017.
	
	\bibitem{2}
	X.~Shao and R.~Zhang, ``6DMA enhanced wireless network with flexible antenna position and rotation: Opportunities and challenges,'' \emph{IEEE Commun. Mag.}, vol.~63, no.~4, pp. 121--128, Apr. 2025.
	
	\bibitem{3}
	K.~Wong and K.~Tong, ``Fluid antenna multiple access,'' \emph{IEEE Trans. Wireless Commun.}, vol.~21, no.~7, pp. 4801--4815, Jul. 2022.
	
	\bibitem{4}
	L.~Zhu, W.~Ma, and R.~Zhang, ``Movable-antenna array enhanced beam forming: Achieving full array gain with null steering,'' \emph{IEEE Commun. Lett.}, vol.~27, no.~12, pp. 3340--3344, Dec. 2023.
	
	\bibitem{1}
	X.~Shao, Q.~Jiang, and R.~Zhang, ``6D movable antenna based on user distribution: Modeling and optimization,'' \emph{IEEE Trans. Wireless Commun.}, vol.~24, no.~1, pp. 355--370, Jan. 2025.
	
	\bibitem{5}
	X.~Shao, R.~Zhang, Q.~Jiang, and R.~Schober, ``6D movable antenna enhanced wireless network via discrete position and rotation optimization,'' \emph{IEEE J. Sel. Areas Commun.}, vol.~43, no.~3, pp. 674--687, Mar. 2025.
	
	\bibitem{7}
	Q.~Jiang, X.~Shao, and R.~Zhang, ``Low-complexity 6DMA rotation and position optimization based on statistical channel information,'' in \emph{Proc. IEEE/CIC Int. Conf. Commun. China (ICCC Workshops)}, Shanghai, China, 2025, pp. 1--6.
	
	\bibitem{8}
	X.~Shao, R.~Zhang, Q.~Jiang, J.~Park, T.~Q.~S.~Quek, and R.~Schober, ``Distributed channel estimation and optimization for 6D movable antenna: Unveiling directional sparsity,'' \emph{IEEE J. Sel. Top. Signal Process.}, vol.~19, no.~2, pp. 349--365, Mar. 2025.
	
	\bibitem{9}
	T.~Ren, X.~Zhang, L.~Zhu, W.~Ma, X.~Gao, and R.~Zhang, ``6-D movable antenna enhanced interference mitigation for cellular-connected UAV communications,'' \emph{IEEE Wireless Commun. Lett.}, vol.~14, no.~6, pp. 1618--1622, Jun. 2025.
	
	\bibitem{10}
	X.~Shao, L.~Hu, Y.~Sun, X.~Li, Y.~Zhang, J.~Ding, X.~Shi, F.~Chen, D.~W.~K.~Ng, and R.~Schober, ``Hybrid near-far field 6D movable antenna design exploiting directional sparsity and deep learning,'' \emph{IEEE Trans. Wireless Commun.}, early access, 2025. 
	
	\bibitem{15}
	3GPP, ``Study on channel model for frequencies from 0.5 to 100 GHz,'' 3rd Generation Partnership Project (3GPP), Tech. Rep. TR 38.901 V14.1.1, Release 14, Aug. 2017. [Online]. Available: \url{http://www.3gpp.org/DynaReport/38901.htm}
	
	\bibitem{28}
	P.~Fan, C.~Feng, Y.~Wang, and N.~Ge, ``Investigation of the time-offset-based QoS support with optical burst switching in WDM networks,'' in \emph{Proc. IEEE Int. Conf. Commun. (ICC)}, New York, NY, USA, 2002.
	
	\bibitem{29}
	P.~Fan and X.~G. Xia, ``Block coded modulation for the reduction of the peak to average power ratio in OFDM systems,'' in \emph{Proc. IEEE Wireless Commun. Netw. Conf. (WCNC)}, New Orleans, LA, USA, 1999.
	
	\bibitem{30}
	Q.~Wu and J.~Zheng, ``Performance modeling and analysis of IEEE 802.11 DCF based fair channel access for vehicle-to-roadside communication in a non-saturated state,'' \emph{Wireless Netw.}, vol.~21, no.~1, pp. 1--11, 2015.
	
	\bibitem{26}
	J.~Zhang, P.~Fan, and K.~B. Letaief, ``Network coding for efficient multicast routing in wireless ad-hoc networks,'' \emph{IEEE Trans. Commun.}, vol.~56, no.~4, pp. 598--607, 2008.
	
	\bibitem{27}
	Z.~Yao, J.~Jiang, P.~Fan, Z.~Cao, and V.~O.~K. Li, ``A neighbor-table-based multipath routing in ad hoc networks,'' in \emph{Proc. 57th IEEE Semiannu. Veh. Technol. Conf. (VTC)}, Jeju, South Korea, 2003.
	
\end{thebibliography}
\end{document}